\documentclass[12pt]{article}
\usepackage{epsfig,amsfonts,amssymb}
\usepackage{hyperref}
\usepackage{cite}
\input epsf.sty
\usepackage{cite}

\def\ZZZ{{\hbox{ Z\kern-1.6mm Z}}}
\def\RRR{{\hbox{ R\kern-2.4mm R}}}
\def\CCC{{\hbox{ C\kern-2.4mm C}}}
\def\zzz{{\hbox{z\kern-1mm z}}}

\newcommand{\qeq}{{\hbox{=\kern-2.3mm ? \kern.5mm }}}
\renewcommand{\qeq}{=}

\newcommand{\MM}{{\cal M}}
\newcommand{\CC}{{\cal C}}

\newcommand{\LL}{{\cal L}}

\newcommand{\wt}{\widetilde}
\newcommand{\wh}{\widehat}

\newcommand{\NN}{{\cal N}}

\newcommand{\crh}{\rho}
\newcommand{\cs}{\sigma}
\newcommand{\cv}{v}

\newcommand{\krh}{\check\rho}
\newcommand{\ks}{\check\sigma}
\newcommand{\kv}{\check v}

\newcommand{\be}{\begin{equation}}
\newcommand{\ee}{\end{equation}}
\newcommand{\ben}{\begin{eqnarray}\displaystyle}
\newcommand{\een}{\end{eqnarray}}

\newcommand{\bea}[1]{\begin{eqnarray}\label{#1} }
\newcommand{\eea}{\end{eqnarray}}

\newcommand{\refb}[1]{(\ref{#1})}

\newcommand{\sectiono}[1]{\section{#1}\setcounter{equation}{0}}

\def\one{{\hbox{ 1\kern-.8mm l}}}
\def\zero{{\hbox{ 0\kern-1.5mm 0}}}

\begin{document}

\baselineskip 24pt

\begin{center}
{\Large \bf Genus Two Surface and Quarter BPS Dyons:}

{\Large \bf The Contour Prescription}

\end{center}

\vskip .6cm
\medskip

\vspace*{4.0ex}

\baselineskip=18pt

\centerline{\large \rm  Shamik Banerjee,  Ashoke Sen
and Yogesh K. Srivastava}

\vspace*{4.0ex}

\centerline{\large \it Harish-Chandra Research Institute}

\centerline{\large \it  Chhatnag Road, Jhusi,
Allahabad 211019, INDIA}

\vspace*{1.0ex}
\centerline{E-mail:  bshamik, sen, yogesh@mri.ernet.in}

\vspace*{5.0ex}

\centerline{\bf Abstract} \bigskip

Following the suggestion of 
hep-th/0506249 and hep-th/0612011, we
represent quarter BPS dyons in $\NN=4$
supersymmetric string theories as string network
configuration and 
explore the role of genus two surfaces in determining the
spectrum of such dyons.
Our analysis leads to the correct contour
prescription for integrating the
partition function to determine the
spectrum in different domains of the moduli space separated by
the walls of marginal stability.

 \vfill \eject

\baselineskip=18pt

\tableofcontents

\sectiono{Introduction} \label{szero}

We now have a good understanding of the spectrum of quarter
BPS dyons in a variety of $\NN=4$ supersymmetric string
theories\cite{9607026,0412287,0505094,
0506249,0508174,0510147,0602254,
0603066,0605210,0607155,0609109,0612011,0702141,
0702150,0705.1433,0705.3874,0706.2363,0707.1563,
0707.3035,0708.1270,0710.4533,0712.0043,
0801.0149,0802.0544,0802.1556,0803.2692,0803.3857,0806.2337}. 
One of the mysteries in these results is the appearance of
modular forms of 
$Sp(2,\ZZZ)$ (or its subgroups)  in the expression
for the dyon partition function. 
Since $Sp(2,\ZZZ)$ is the modular group of genus two Riemann
surfaces, one might expect that genus two surfaces would play
a role in determining the dyon spectrum. The counting that
leads to the result however does not explicitly make use of
genus two Riemann surfaces\cite{0505094,0605210}.
A possible explanation for the role of genus two
surfaces has been suggested in 
\cite{0506249,0612011} by representing the quarter BPS dyon as
a string network configuration\cite{9704170,9710116,9711130} 
in type IIB string theory on
$K3\times T^2$ and then relating the associated partition
function via duality to a configuration
of euclidean M5-branes wrapped on $K3$ times a genus two 
Riemann surface.

Another mystery in this subject
is the prescription for computing the spectrum
in different domains in the moduli space separated by walls
of marginal stability. Naively one would expect that since the
dyon spectrum jumps discontinuously across a wall of
marginal 
stability\cite{0005049,0010222,0101135,0206072,0304094,0702146}, 
the partition
function computed in different domains will be different. 
Instead
one finds that as an analytic function of the chemical potentials
the partition functions in different domains are 
identical\cite{0605210,0609109,0702141}.
However, in order
to extract the degeneracies from the partition function in different
domains in the moduli space, one needs to choose different contours
in the space of complex chemical potentials along which we carry
out the 
Fourier integral of the
partition function. A specific set of rules relating the domains in
the moduli space and the integration contour in the space of chemical
potentials 
have 
been given in \cite{0702141,0706.2363}. 
The original prescription of \cite{0702141} restricts the location of the
integration contour to be inside a certain region depending on the
domain in the moduli space where we want to compute the
degeneracy. The value of the integral is independent of the choice
of contour as long as the contour lies within this region. 
This
prescription arises from explicit counting of states of quarter
BPS dyons\cite{0605210,0609109} 
and the requirement of S-duality invariance\cite{0702141,0702150}.
Ref.\cite{0706.2363}
proposed a definite choice of contour corresponding to a specific
point in the moduli space. While this is consistent with the
prescription of \cite{0702141}, 
currently we do not have any understanding of the
physical origin of this prescription.
Our main goal in this paper will be to
derive this prescription using the relation between the dyon
spectrum and genus two surfaces, 
and in that process make the
approach advocated  in \cite{0506249,0612011} a little more
precise. For simplicity we work with the specific example of
heterotic string theory on $T^6$, but the results should be easily
generalizable to other $\NN=4$ supersymmetric string theories.

The rest of the paper is organised as follows. In \S\ref{sintro}
we review some of the necessary background material and summarize
our main results. In \S\ref{sfivebrane} we describe a specific set
of quarter BPS dyon configurations in heterotic string theory
on $T^6$, introduce
(real) chemical potentials dual  to  appropriate charges
and relate these chemical
potentials to background values of appropriate components
of 2-form 
fields in the theory. We then show how the dyon partition
function associated with these states automatically complexifies the
chemical potentials and leads to the
correct choice of the integration contour in the space of complex
chemical potentials.
In \S\ref{sgeometry}
we use the strategy of \cite{0506249,0612011} to relate the
original partition function to that of an M5-brane on $K3$
times a genus two surface
and show that the moduli of the associated genus two surface
are given precisely by the complexified chemical potentials
which arise in the analysis of \S\ref{sfivebrane}. However
our analysis does not lead to a foolproof derivation of the actual
partition function. Some of the problems were already discussed
in \cite{0612011}; we discuss some additional subtleties
in \S\ref{subtle}.

Finally we would like to note that the analysis of this paper can
in principle be extended to 1/8 BPS states in type IIB string
theory on $T^4\times T^2$ by representing these states as a
network of $(p,q)$ 5-branes along $T^4$ 
times cycles of $T^2$.
This suggests that this 
partition function
can also be represented
as an appropriate quantity associated with a genus two Riemann
surface.

\sectiono{Background and Summary of Results} \label{sintro}

In this section we shall review some material 
which will be needed for our analysis, and then summarize
our results.
We shall restrict our analysis to the simplest $\NN=4$
supersymmetric string theory, namely
heterotic string theory compactified on $T^6$ or equivalently
type IIA or IIB string theory compactified on $K3\times T^2$.
This theory has 28 $U(1)$ gauge fields and as a result a dyon
is characterized by a pair of 28 dimensional vectors $(Q,P)$
labelling
electric and magnetic charges. The T-duality group of the theory
is a discrete subgroup of $O(6,22)$ 
with $Q$ and $P$ transforming in the vector representation of
$O(6,22)$. 
We denote by $L$ the
signature $(6,22)$ matrix that remains invariant under the
$O(6,22)$ transformation and define
\be \label{ei1}
Q^2=Q^TLQ, \qquad P^2=P^TLP, \qquad Q\cdot P
= Q^TLP\, .
\ee
These are the only independent 
invariants of the continuous $O(6,22)$
group which can be constructed out of $Q$ and $P$.
However since only a discrete subgroup of $O(6,22)$ 
is a
symmetry of string theory, there are other invariants of this
discrete group on which physical quantities can depend. 
$\gcd(Q\wedge P)$ is one such 
invariant\cite{0702150} 
We shall restrict to a subset of dyons for which
\be \label{ei2}
\gcd(Q\wedge P) = 1\, .
\ee
One can show that within this subclass 
$(Q^2,P^2,Q\cdot P)$ are the
complete set of T-duality invariants, \i.e.\ two dyon charges
satisfying \refb{ei2} and having same values of $Q^2$, $P^2$ and
$Q\cdot P$ can be related by a T-duality 
transformation\cite{0712.0043}.

We shall denote by $d(Q,P)$ the sixth
helicity trace index $B_6$ of 
dyons of charge $(Q,P)$\cite{9708062}. 
This effectively counts the number of
quarter BPS supermultiplets carrying charge $(Q,P)$, with sign
$+1$ ($-1$) if the average helicity of all the states in the
supermultiplet is an integer (integer plus half). This is a protected
index and hence is not expected to change under a continuous
variation of the moduli of the theory. However there are
walls of marginal stability on which a quarter BPS dyon can break
up into a pair of half-BPS dyons, and as we cross such a wall in
the moduli space the index $d(Q,P)$ can jump. 
A simple way to label the walls of marginal stability is as
follows. One can show that for dyons carrying charge
vector of the form given in \refb{ei2} the
decay of a quarter BPS
dyon into half BPS dyons can take place on a codimension
one subspace only if the decay is of the form\cite{0707.1563}:
\be \label{ef1}
(Q,P) \to (\alpha Q + \beta P, \gamma Q+\delta P) 
+ (\delta Q - \beta P, -\gamma Q + \alpha P), \quad
\alpha+\delta=1, \quad \alpha\delta=\beta\gamma \, ,
\ee
where $(\alpha,\beta,\gamma,\delta)$ are integers.
Given $(\alpha,\beta,\gamma, \delta)$, the corresponding 
wall in the moduli space may be determined by solving 
the equation
\be \label{ef2}
m(Q,P) = m(\alpha Q + \beta P, \gamma Q+\delta P) 
+ m(\delta Q - \beta P, -\gamma Q + \alpha P)\, ,
\ee
where $m(Q,P)$ denotes the BPS mass of the state of charge
$(Q,P)$. Thus for given $(Q,P)$,
specifying $(\alpha,\beta,\gamma,\delta)$
determines the wall uniquely, and we can label a wall by the
set of integers $(\alpha,\beta,\gamma,\delta)$. 

For a given charge $(Q,P)$,
these walls of marginal stability divide the moduli space
of vacua into different domains. 
A domain
bounded by a set of walls can then be specified by giving
the values of $(\alpha,\beta,\gamma,\delta)$ for each of the
walls bordering the domain. We shall denote
the collection of $(\alpha,\beta,\gamma,\delta)$ labelling
a domain by $\vec c$. Thus $d(Q,P)$ depends not only on
the T-duality invariants $Q^2$, $P^2$ and $Q\cdot P$, but
also on the domain label $\vec c$. Let us denote this
function by $f(Q^2/2,P^2/2, Q\cdot P; \vec c)$, and define the
partition function:
\be \label{ef3}
Z(\rho,\sigma,v; \vec c) \equiv
\sum_{Q^2,P^2,Q\cdot P} \, (-1)^{Q\cdot P+1} 
\, \exp\left\{2\pi i \left(\sigma \, {Q^2\over 2}
+ \rho\, {P^2\over 2} + v\, Q\cdot P\right)\right\}\, 
f(Q^2/2,P^2/2, Q\cdot P; \vec c) \, .
\ee
Since the dyon degeneracy function $f$ grows rapidly 
for large charges, the sum given above is not convergent
for real $\rho$, $\sigma$, $v$, -- we need to define it in the
complex $(\rho,\sigma,v)$ space. 
Let us take
\be \label{ef4a}
\rho=\rho_1+i\rho_2, \qquad \sigma=\sigma_1+i\sigma_2, \qquad
v = v_1+iv_2\, .
\ee
In turns out that the sum converges
only inside a certain region in the $(\rho_2,\sigma_2,v_2)$
space. We shall refer to these regions as `chambers', -- whereas
the word `domain' will be reserved for the regions in the
moduli space of vacua bounded by walls of marginal stability.
This chamber depends on the domain label $\vec c$ 
as follows.
Suppose $(\alpha,\beta,\gamma, \delta)$ are the parameters
associated with a particular wall. We associate with this a
plane in the $(\rho_2,\sigma_2,v_2)$ space given by
\be \label{ef5}
\crh_2 \gamma - \cs_2 \beta+ \cv_2 (\alpha-\delta) = 0\, .
\ee
Now consider the collection of such planes for all 
$(\alpha,\beta,\gamma,\delta)$ associated with the walls of
a particular
domain. These form the boundary of 
a cone in the $(\rho_2,\sigma_2,v_2)$
space with its vertex at the origin. 
The chamber in the $(\rho_2,\sigma_2,v_2)$ space
where the sum
\refb{ef3} is convergent consists of points inside this chamber
lying sufficiently far away from the origin.
In other words if we
pick any point
$(a_1,a_2,a_3)$ in the interior of this cone and choose
\be \label{ef6}
(\rho_2,\sigma_2,v_2) = 
\Lambda (a_1,a_2,a_3),  
\ee
then for sufficiently large $\Lambda$ the sum converges.
Once the partition function is defined this way inside a
chamber, one can extend
it to other regions in the complex $(\rho,\sigma,v)$ space
via analytic continuation. It turns out that the function defined
this way is
independent of the choice of $\vec c$ and is given by the inverse
of the Igusa cusp form $\Phi_{10}(\rho,\sigma,v)$:
\be \label{epart}
Z(\rho,\sigma,v;\vec c) = {1\over \Phi_{10}(\rho,\sigma,v)}\, .
\ee
This allows us to invert \refb{ef3} to write
\be \label{ef3a}
d(Q,P) = (-1)^{Q\cdot P+1} \int_{\CC(\vec c)}
 d\crh
d\cs  d\cv \,
e^{-i\pi (\cs Q^2 + \crh P^2 
+ 2 \cv Q\cdot P)} \, 
{1\over \Phi_{10}(\crh, \cs, \cv)}\, ,
\ee
where the choice
of the contour $\CC(\vec c)$ is given by
\be \label{ef4}
0\le \rho_1\le 1, \quad 0\le \sigma_1\le 1, \quad 0\le v_1\le 1,
\quad \rho_2=M_1, \quad \sigma_2=M_2, \quad v_2=M_3\, ,
\ee
$M_1$, $M_2$, $M_3$ being constants lying inside the
chamber where the original sum is convergent.
The result \refb{ef3a} was derived by working in a given domain
of the moduli space where the type IIB string coupling is 
weak\cite{0605210,0609109}, and
then extending the result to other domains by S-duality 
transformation\cite{0702141,0702150}.

A simple prescription for the choice of $(\rho_2, \sigma_2, v_2)$
that satisfies the requirement of lying inside a given chamber
when the moduli lie inside a given domain
was given in \cite{0706.2363}. Heterotic string theory
on $T^6$ contains a complex scalar modulus $\tau$ labelling
the axion-dilaton field and another set of 132 real moduli
labelling the coset $O(6,22)/(O(6)\times O(22))$. The latter
are
parametrized by a symmetric $28\times 28$ matrix $M$
satisfying $M^TLM=L$. We define
\be \label{edefcrint}
Q_R^2 = 
Q^T (M+L)Q, \quad P_R^2=P^T (M+L)P, \quad
Q_R\cdot P_R =  Q^T(M+L)P\, .
\ee
Then if we choose\footnote{Throughout this paper we shall use
the convention of \cite{0802.0544}.}
\ben \label{eimdefpre}
\rho_2 &=& \Lambda \,
 \left\{ {|\tau|^2 \over \tau_2} + {Q_R^2\over
 \sqrt{Q_R^2 P_R^2 - (Q_R\cdot P_R)^2}}\right\}\, ,
  \nonumber \\
 \sigma_2 &=& \Lambda  \,
 \left\{{1\over \tau_2}
+ {P_R^2\over
 \sqrt{Q_R^2 P_R^2 - (Q_R\cdot P_R)^2}}\right\}\, ,
  \nonumber \\
v_2 &=& -\Lambda  \,
\left\{ {\tau_1 \over \tau_2} + {Q_R\cdot P_R
 \over
 \sqrt{Q_R^2 P_R^2 - (Q_R\cdot P_R)^2}}\right\}\, ,
 \een
 then for sufficiently large $\Lambda$, $(\rho_2,\sigma_2,v_2)$ 
 automatically lie inside the correct chamber associated
 with the domain in which 
the point $(\tau,M)$ lie.
This formula picks a given ray inside the cone bounded by the
surfaces \refb{ef5}.  As far as the prescription for the contour
is concerned, such precise specification is not
necessary; any other ray inside the cone would have been an
equally good choice. Nevertheless this formula has some remarkable
properties. It correctly takes us from one chamber to another
as the moduli cross a wall of marginal stability. Furthermore
this formula is S-duality covariant; if we pick another point
in the moduli space related to the original one by an S-duality
transformation, the $\rho_2$, $\sigma_2$, $v_2$ given in
\refb{eimdefpre} transform correctly so as to preserve the
exponent in \refb{ef3a}. This makes one feel that there must
be some deeper origin of this formula that would also naturally
explain the correlation between the chambers in the
$(\rho_2,\sigma_2,v_2)$ space and domains in the
moduli space without having to make
use of S-duality transformation.

However already at this stage we can anticipate a possible
difficulty in deriving \refb{eimdefpre}. In defining the
dyon partition function via \refb{ef3} we need to sum over
different charges at {\it fixed values of $(\rho,\sigma,v)$}. 
On the other hand \refb{eimdefpre} determines the imaginary
parts of $(\rho,\sigma,v)$ as a function of the charges and moduli.
Thus if we keep the moduli fixed, it would seem that we need to
keep changing the imaginary parts of $(\rho,\sigma,v)$ as we
sum over different charges.\footnote{In contrast if we keep the
label $\vec c=\{(\alpha_i,\beta_i,\gamma_i, \delta_i)\}$ fixed, then
there is no problem of this kind since the restriction on
$(\rho_2,\sigma_2,v_2)$ to lie inside the cone bounded by the
surfaces \refb{ef5} is independent of the charges.} 
How can we satisfy these two
contradictory requirements? We shall solve this 
problem by working in
a specific corner of the moduli space and with
a specific family of dyon
charges  such that as we vary the charges to generate different
$(Q^2,P^2,Q\cdot P)$, the values of 
$(\rho_2,\sigma_2,v_2)$ computed from \refb{eimdefpre}
remain unchanged. In this case we do not
have any difficulty in defining the dyon partition function
at fixed values of the moduli.

Like the contour prescription \refb{eimdefpre}, the appearance of 
the Igusa cusp
form $\Phi_{10}$ in the expression for the dyon partition function
is also quite mysterious.
The same cusp form 
also appears in the expression for the two loop
partition function of the bosonic string theory. This would
lead one to suspect that there is an underlying genus two
surface behind the formula given in \refb{ef3a}.
There is however 
no sign of such a genus two surface in the counting that
leads to \refb{ef3a}; the final result just happens to have this
specific form. 
In a pioneering work, Gaiotto\cite{0506249} and later
Dabholkar and Gaiotto\cite{0612011} suggested a possible
origin of this genus two surface from a different viewpoint. 
The main idea of 
\cite{0506249, 0612011} was to represent the quarter 
BPS configuration
in heterotic string theory on $T^6$ as a network of strings in a
dual type IIB string theory on $K3\times T^2$. The strings are obtained
by wrapping $(p,q)$ 
five-branes  on $K3$, and the network of such strings lie
along the plane of the $T^2$. The partition function of such strings
(with $(-1)^F$ inserted in the trace and the integration over the
fermionic and the bosonic zero modes associated with the
center of mass degrees of freedom factored out) 
can be represented as a
path integral over an euclidean type IIB string theory with
periodic boundary condition along the thermal circle. By making
a T-duality transformation along the thermal circle and then
identifying the resulting type IIA string theory as M-theory on
another circle, the partition function can be regarded as that
of an euclidean M5-brane wrapped on $K3$ times a genus two
surface embedded in $T^4$. This genus two surface was identified
as the origin of the $\Phi_{10}$ in the partition function
of quarter BPS states.

Our main purpose in this paper is to make this procedure a little
more precise and in that process recover the correct prescription
for the integration contour as given in \refb{eimdefpre}. 
For this we
begin with a configuration in
IIB on $K3\times T^2$ with
D5 and NS5-branes wrapped on $K3$ times 
two different cycles of $T^2$
and also D strings and fundamental strings wrapped on various
cycles of $T^2$. 
For fixed D5 and NS5-brane charges 
$Q^2$, $P^2$ and $Q\cdot P$ are
given by appropriate linear combinations of the D-string and
fundamental string charges, and hence 
the chemical potentials dual to $Q^2/2$, $P^2/2$ and 
$Q\cdot P$
can be interpreted as background values of the 2-form fields
with one leg along the time direction and another leg along the
cycles of $T^2$. 
We denote these background fields by $\sigma_1$,
$\rho_1$  and $v_1$ respectively.
We then euclideanize the time circle 
and compactify it on a circle of period $2\pi\beta$
as in \cite{0506249, 0612011}, with perodic boundary condition
on the fermions.
The euclidean path integral in the presence of such a
background may be  represented as a trace of $(-1)^F e^{-2\pi 
\beta H}$
with an extra insertion of the 
$\exp\left[ 2\pi i \left( \rho_1 {P^2\over 2} 
+ \sigma_1 {Q^2\over 2}
 + v_1 Q\cdot P \right) \right] $ representing the effect of
 the background 2-form fields. Furthermore we do not need
 any extra damping factor for regulating the trace;  
 the damping is provided
 by the $\exp(-2\pi \beta \, m(Q,P))$ term that appears
 naturally in the trace, $m(Q,P)$ being
 the mass of the BPS state of charge $(Q,P)$. 
By expanding $\exp(-2\pi \beta \, m(Q,P))$ 
in appropriate limit where the 5-branes give the dominant
contribution to $m(Q,P)$ we
find that it effectively provides us with a damping factor
of $\exp\left\{ -2\pi \left(  \rho_2 {P^2\over 2} 
+ \sigma_2 {Q^2\over 2}
 + v_2 Q\cdot P \right) \right\}$ with $(\rho_2,\sigma_2,v_2)$
 given in \refb{eimdefpre}. Thus this procedure automatically leads
 to the choice of $(\rho_2, \sigma_2, v_2)$ that makes the
 partition function convergent.
 
  We then go ahead and follow the prescription of 
 \cite{0506249, 0612011} to map the euclidean IIB theory to
 euclidean M-theory on $K3\times T^4$ and the string network
 configuration to a configuration of euclidean
 M5-brane wrapped on
 $K3$ times a genus two surface embedded in $T^4$. 
 Standard duality transformation
 laws determine the geometry of the final $T^4$. 
 This in turn allows us to find the moduli
 of the genus two surface by requiring that the surface is
 holomorphically embedded in $T^4$. The result is that the
 period matrix of the genus two surface is given by
 \be \label{eff1}
 \Omega = \pmatrix{\sigma_1+i\sigma_2 & v_1+iv_2\cr 
 v_1+iv_2 & \rho_1+i\rho_2}\, ,
 \ee
 with $(\rho_1,\sigma_1,v_1)$ determined by the background
 2-form fields in the original theory, and 
 $(\rho_2,\sigma_2,v_2)$ given by \refb{eimdefpre}.
 This allows us the relate the variables $(\rho,\sigma,v)$
 appearing in the definition of the partition function
 to the moduli of genus two surfaces.
 
While our analysis tells us that the partition function of
quarter BPS states is given by an appropriate partition
function on the genus two surface with moduli $(\rho,\sigma,v)$,
we have not explicily computed the partition function and shown
that it is given by the inverse of the Igusa cusp form. 
Refs.\cite{0506249, 0612011} 
already made some progress in this direction,
but some subtle points involving fermion zero modes are
yet to be sorted out. We point out in \S\ref{subtle} some additional
issues
in the analysis of the partition function. We hope to
return to these points in future.

\sectiono{Dyon Partition Function from 5-brane 1-brane System} 
\label{sfivebrane}

We begin with type IIB string theory compactified 
on $K3\times S^1\times
\wt S^1$. 
By making a T-duality transformation
on the circle $\wt S^1$ to map this to
type IIA string theory on $K3\times S^1\times \wh S^1$ and then
using the duality between type IIA string theory on K3 and heterotic
string theory on $T^4$, we can map this theory to heterotic
string theory on $T^4\times S^1\times \wh S^1$.\footnote{This
is a slightly different set of duality transformations compared to
those used {\it e.g.} in \cite{0605210}.}
Under this duality
states in the IIB theory
carrying winding charge along $S^1$ get mapped 
to magnetically charged states in the heterotic string theory
and the states carrying winding charge
along $\wt S^1$ get mapped to electrically charged states
in the heterotic string theory. Furthermore the complex structure
modulus $-\tau_1+i\tau_2$ 
of the $S^1\times \wt S^1$ torus, with the
$\wt S^1$ regarded as the $a$-cycle, gets mapped to the axion-dilaton
modulus $\tau\equiv \tau_1+i\tau_2$  
of the dual heterotic string. The other moduli of the
IIB theory get mapped to the $O(6,22)/ (O(6)\times O(22))$
moduli $M$ of the heterotic string theory, but we shall not need to
know the explicit form of this map.

Consider a state in the IIB description containing a $(p_1,q_1)$
5-brane wrapped on $K3\times \wt S^1$ and a $(p_2,q_2)$ 5-brane
wrapped on $K3\times  S^1$. If we regard the K3-wrapped 5-branes
as strings then such a configuration forms a network of strings on
$S^1\times \wt S^1$\cite{0506249}, and
the BPS mass of this object, measured in type IIB metric,
is given by\cite{9711130}:\footnote{In our convention
a $(p,q)$ five brane carries $p$ units of D5-brane charge and
$q$ units of NS 5-brane charge. Ref.\cite{9711130} gives the
mass formula for the D-string - fundamental string
system in the 
ten dimensional 
canonical metric; we have rescaled it by
an appropriate factor to convert it
to type IIB metric and also to take into account extra factors
of string coupling and the volume of $K3$ which appear in the
expression for the $(p,q)$
five brane mass compared to the $(p,q)$ string
mass.}
\be \label{e12}
m^2_{IIB} = A \, (V_{K3})^2 \, \lambda_2^3\, 
\pmatrix{p_1 & q_1 & p_2 & q_2} (M_0\pm L_0)
\pmatrix{p_1\cr q_1 \cr p_2\cr q_2}
\ee
where $A$ denotes the area of $S^1\times \wt S^1$ and $V_{K3}$
is the volume of $K3$ measured in the type IIB metric, 
and 
\be \label{e13}
L_0 \equiv \pmatrix{0 & \LL\cr -\LL & 0}, \qquad
\LL \equiv \pmatrix{0 & 1\cr -1 & 0}\, ,
\ee
\be \label{e14}
M_0 \equiv {1\over \tau_2}
 \pmatrix{\MM & -\tau_1 \MM \cr -\tau_1 \MM  
& |\tau|^2 \MM}, \qquad \MM \equiv {1\over \lambda_2} \, 
\pmatrix{1 & -\lambda_1 \cr -\lambda_1 & |\lambda|^2}\, .  
\ee
Here $\lambda=\lambda_1+i\lambda_2$ denotes the axion-dilaton
modulus of the ten dimensional type IIB string theory. 
The sign in front of $L_0$ in \refb{e12} is to be chosen so
that the contribution of this term to mass$^2$ is positive.
On the other hand for most general set of charges $(Q,P)$ in the
heterotic description, 
the BPS mass formula
measured in the four dimensional canonical metric takes the form
\be \label{e15pre}
m_{can}(Q,P)^2 = \left\{{1\over \tau_2}  |Q_R - \tau P_R|^2 
+ 2 \, \sqrt{Q_R^2 P_R^2 - (Q_R\cdot P_R)^2}\right\}\, ,
\ee
where $Q_R^2$, $P_R^2$ and $Q_R\cdot P_R$ have been defined
in \refb{edefcrint}.
Since the canonical four dimensional
metric $g_{\mu\nu}$
and the type IIB metric $g^{IIB}_{\mu\nu}$ are related by
\be \label{ecan}
g_{\mu\nu} = V_{K3} \, A\, \lambda_2^2 \, g^{IIB}_{\mu\nu}\, ,
\ee
the BPS mass$^2$ measured in the type IIB metric takes the
form
\be \label{e15}
m(Q,P)^2 = V_{K3} \, A\, \lambda_2^2 \, 
\left\{{1\over \tau_2}  |Q_R - \tau P_R|^2 
+ 2 \, \sqrt{Q_R^2 P_R^2 - (Q_R\cdot P_R)^2}\right\}\, .
\ee

\begin{figure}
\begin{center}
\hbox{\qquad \qquad 
\qquad \qquad \qquad \qquad \epsfysize=4cm \epsfbox{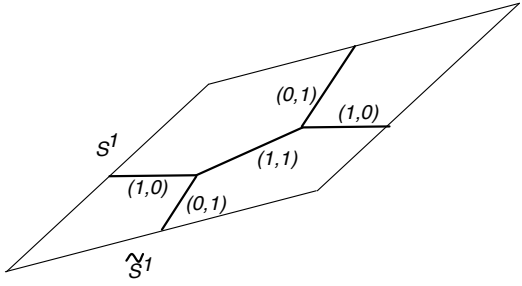}
}
\end{center}
\caption{The string network configuration.} \label{f1}
\end{figure}

We now consider a configuration 
consisting of a $(1,0)$ 5-brane, \i.e.\
a D5-brane, wrapped on $K3\times \wt S^1$ and a (0,1) 5-brane,
\i.e.\ an NS5-brane, wrapped on $K3\times  S^1$. Thus we have
$(p_1,q_1)=(1,0)$ and $(p_2,q_2)=(0,1)$. Since the
(1,0) brane wraps $\wt S^1$ it represents 
an electric charge vector $Q_0$
in the dual heterotic string theory.
On the other hand the $(0,1)$ brane being wrapped on $S^1$
represents a magnetic charge $P_0$. 
The combined configuration may be represented as a string
network on $S^1\times \wt S^1$ as shown in Fig.~\ref{f1}.
 Our first task will be to express the combinations $Q_{0R}^2$,
 $P_{0R}^2$ and $Q_{0R}\cdot P_{0R}$
 in terms of type IIB variables. This is done by comparing the
 BPS mass formul\ae\ \refb{e12}, \refb{e15} applied to
 the charge vectors $(Q_0,0)$, $(0,P_0)$ and $(Q_0,P_0)$.  
 We get\footnote{A D5-brane wrapped on $K3$ carries 
 $1$  
 unit
 of D1-brane charge and an NS 5-brane wrapped on $K3$ carries
 $1$ unit of fundamental string charge. In \refb{e21}, \refb{e22}
 we have assumed that these 1-brane charges have been neutralized
 by appropriate configuration of 
 D-string and fundamental string.}
 \ben \label{e21}
 m(Q_0,0)^2 &=& V_{K3} \, A\, \lambda_2^2 \, 
 {Q_{0R}^2\over \tau_2} = A \, (V_{K3})^2
 \, \lambda_2^3\,
 {1 \over \lambda_2
 \tau_2 }\, ,
 \nonumber \\
 m(0,P_0)^2 &=& V_{K3} \, A\, \lambda_2^2 \, 
 {P_{0R}^2 |\tau|^2 \over \tau_2} = A \, (V_{K3})^2
 \, \lambda_2^3\, 
 {|\tau|^2|\lambda|^2 \over \lambda_2 \tau_2}\, , \nonumber \\
m(Q_0,P_0)^2 &=& V_{K3} \, A\, \lambda_2^2 \, 
\left\{{1\over \tau_2}  |Q_{0R} - \tau P_{0R}|^2 
+ 2 \, \sqrt{Q_{0R}^2 P_{0R}^2 - (Q_{0R}\cdot P_{0R})^2}\right\} \nonumber \\
&=&   A \, (V_{K3})^2\, \lambda_2^3\, \left[
{1 \over \lambda_2\tau_2 } + {|\tau|^2|\lambda|^2\over 
\lambda_2\tau_2} + 2{\tau_1 
\lambda_1 \over
\lambda_2\tau_2} 
+ 2\right]\, . \nonumber \\
 \een
 This gives
 \be \label{e22}
 Q_{0R}^2 = V_{K3}, 
 \qquad P_{0R}^2 = V_{K3}\,  
 {|\lambda|^2 }, \qquad Q_{0R}\cdot P_{0R}
 = -V_{K3}\,  {\lambda_1 }\, .
 \ee
 
We now
add to the previous system 
$n_1$ units of fundamental string  
charge wrapped on
$\wt S^1$, $n_2$ units of fundamental string charge wrapped on
$S^1$, $m_1$ units of D-string charge wrapped on $\wt S^1$ and
$m_2$ units of D-string charge wrapped on $S^1$.
These can be regarded as excitations of the original D5-brane -
NS5-brane system involving small instantons and world-volume
electric fields.  Unbroken supersymmetry imposes the constraints
$m_1\ge -1$, $n_2\ge -1$ after taking into account the induced
D-string and the fundamental string charges on the $K3$ wrapped
D5 and NS5-branes.
Then by following the standard
duality chain to map these charges into the heterotic description,
we get
\be \label{e1}
Q^2 \equiv Q^TLQ= 2 m_1, \quad P^2\equiv P^TLP
 = 2 n_2, \quad 
Q\cdot P \equiv Q^TLP= m_2+n_1\, .
\ee
The expression for the mass $m(Q,P)$ gets modified in the
presence of these charges. We shall assume that $V_{K3}$ is
large so that the 5-branes still give the dominant contribution
to $m(Q,P)$ and compute the first order correction to $m(Q,P)$
due to the D1 brane and fundamental string charges. It is easy to
see that in eq.\refb{e15}
this contribution comes from the correction to $Q_R^2$,
$P_R^2$ and $Q_R\cdot P_R$ given in \refb{edefcrint}
from the
$Q^2=2m_1$, $P^2=2n_2$ and $Q\cdot P=m_2+n_1$ terms
respectively, -- corrections to $Q^TMQ$, $P^TMP$ and
$Q^TMP$ terms are suppressed by further powers of 
$V_{K3}$.\footnote{This requires 
taking $V_{K3}\to\infty$ limit in such a way that the off-diagonal
components of $M$ which couple the wrapped 5-brane charges
to the wrapped 1-brane charges vanish in this limit.}
This gives
\ben \label{ede1}
\delta m(Q,P)  &=& {1\over m(Q,P)} \, 
V_{K3}\, A\, \lambda_2^2\, 
\left[ {Q^2\over 2} \left\{{1\over \tau_2}
+ {P_{R}^2\over
 \sqrt{Q_{R}^2 P_{R}^2 - (Q_{R}\cdot P_{R})^2}}\right\}
 \right. \nonumber \\ &&
 \left. + {P^2\over 2} 
 \left\{ {|\tau|^2 \over \tau_2} + {Q_{R}^2\over
 \sqrt{Q_{R}^2 P_{R}^2 - (Q_{R}\cdot P_{R})^2}}\right\}  - Q\cdot P 
 \left\{ {\tau_1 \over \tau_2} + {Q_{R}\cdot P_{R}
 \over
 \sqrt{Q_{R}^2 P_{R}^2 - (Q_{R}\cdot P_{R})^2}}\right\} \right]\, 
 \nonumber \\
&=& A^{1/2} \, \lambda_2^{1/2}\,  \left[
{1 \over \lambda_2\tau_2 } + {|\tau|^2|\lambda|^2\over 
\lambda_2\tau_2} + 2{\tau_1 
\lambda_1 \over
\lambda_2\tau_2} 
+ 2\right]^{-1/2} \, \nonumber \\ &&
\left[ {Q^2\over 2} \left\{{1\over \tau_2}
+ {P_{R}^2\over
 \sqrt{Q_{R}^2 P_{R}^2 - (Q_{R}\cdot P_{R})^2}}\right\}
 + {P^2\over 2} 
 \left\{ {|\tau|^2 \over \tau_2} + {Q_{R}^2\over
 \sqrt{Q_{R}^2 P_{R}^2 - (Q_{R}\cdot P_{R})^2}}\right\}  
 \right. \nonumber \\  && \left. - Q\cdot P 
 \left\{ {\tau_1 \over \tau_2} + {Q_{R}\cdot P_{R}
 \over
 \sqrt{Q_{R}^2 P_{R}^2 - (Q_{R}\cdot P_{R})^2}}\right\} \right]\, 
 \nonumber \\
  \een
 where in the last step we have used the leading order
 expression for $m(Q,P)$ given in the last line of \refb{e21}.
 Note that on the right hand side of this equation we have
 used the arguments $Q$, $P$ instead of $Q_0$, $P_0$.
 Since the difference between $(Q_R^2,P_R^2,Q_R\cdot P_R)$ 
 and $(Q_{0R}^2,P_{0R}^2,Q_{0R}\cdot P_{0R})$ is already
 of the first order, the error due to the replacement of $(Q_0,P_0)$
 by $(Q,P)$ on the right hand side is of higher order.
We shall work in the limit $V_{K3}\to \infty$ at fixed
$\lambda$, $\tau$ and 
$A$; in this limit
\refb{ede1} is the exact expression for $\delta m(Q,P)$.

We 
 denote by $y$ and $\wt y$ the coordinates along
 $S^1$ and $\wt S^1$ respectively, and by $t$ the time coordinate.
 Let $C^{(2)}$ and $B^{(2)}$ denote
 the RR and NSNS 2-form fields.
 We now
 make a Wick rotation
 $t\to - i\tau$, compactify the $\tau$ coordinate on a circle
 of period $2\pi\, \beta$ with periodic boundary condition
 along the circle, and switch on background values
 of $C^{(2)}_{\tau y}$, $C^{(2)}_{\tau\wt y}$, $B^{(2)}_{\tau y}$
 and $B^{(2)}_{\tau \wt y}$ of the form
 \be \label{en1}
 C^{(2)}_{\tau y}=B^{(2)}_{\tau\wt y} = 
 v_1, \quad C^{(2)}_{\tau\wt y}
 = \sigma_1, \quad B^{(2)}_{\tau y} = \rho_1\, ,
 \ee
 all normalized so that $\rho_1$, $\sigma_1$ and $v_1$ 
 have period 1.
Since 
the background fields $C^{(2)}_{\tau \wt y}$, 
$C^{(2)}_{\tau y}$, $B^{(2)}_{\tau \wt y}$
 and $B^{(2)}_{\tau  y}$ couple to the charges
 $m_1$, $m_2$, $n_1$ and $n_2$ respectively,
 the presence of the background
 \refb{en1}
corresponds to
 inserting a factor of\footnote{In our convention   
$(C^{(2)}_{MN},B^{(2)}_{MN})$ and $(-n_i, m_i)$
transform as $SL(2,R)$ doublets. Thus the exponent given in
\refb{e43} is S-duality invariant.}
 \be \label{e43}
\exp\left[ 2\pi i \left( C^{(2)}_{\tau\wt y} m_1 + C^{(2)}_{\tau y} m_2
+ B^{(2)}_{\tau\wt y} n_1 + B^{(2)}_{\tau y} n_2\right)\right]
=
\exp\left[ 2\pi i \left( \rho_1 {P^2\over 2} + \sigma_1 {Q^2\over 2}
 + v_1 Q\cdot P \right) \right] \, ,
 \ee
 into the functional integral. 
 In the Hamiltonian formulation this functional integral may
 be represented as a trace with an additional insertion of
 $e^{-2\pi \beta H} (-1)^{F}$.\footnote{The insertion
 of $(-1)^F$ into the trace reflects the effect of putting periodic
 boundary condition on the fermions
 along the $\tau$ direction. We shall assume as in
 \cite{0506249,0612011,0702146} 
 that the trace over the center of mass degrees
 of freedom and their fermionic superpartners have been factorized;
 otherwise trace over these additional zero modes will make the
 result vanish, and we need to insert six powers of helicity into the
 trace to get a non-zero answer.}
 Identifying $H$ as the mass of the brane configuration
 and removing the contribution to $H$ from the leading 
 term 
 $m(Q_0,P_0)$ that does not depend on $m_i$, $n_i$, we see that the
 path integral computes the quantity
 \ben \label{epath2}
&& Tr\left[ (-1)^F
 \exp\left\{ 2\pi i \left( \rho_1 {P^2\over 2} + \sigma_1 {Q^2\over 2}
 + v_1 Q\cdot P \right) \right\}
 \exp\left\{-2\pi \beta \delta m(Q,P)\right\} \right] \nonumber \\
&=& Tr\left[ (-1)^F
 \exp\left\{ 2\pi i \left( \rho_1 {P^2\over 2} + \sigma_1 {Q^2\over 2}
 + v_1 Q\cdot P \right) \right\}
\exp\left\{ -2\pi \left(  \rho_2 {P^2\over 2} + \sigma_2 {Q^2\over 2}
 + v_2 \, Q\cdot P \right) \right\}\right]\nonumber \\
  \een
 where
 \ben \label{eimdef}
 \rho_2 &=& \Lambda \,
 \left\{ {|\tau|^2 \over \tau_2} + {Q_R^2\over
 \sqrt{Q_R^2 P_R^2 - (Q_R\cdot P_R)^2}}\right\} \nonumber \\
 \sigma_2 &=& \Lambda  \,
 \left\{{1\over \tau_2}
+ {P_R^2\over
 \sqrt{Q_R^2 P_R^2 - (Q_R\cdot P_R)^2}}\right\} \nonumber \\
v_2 &=& -\Lambda  \,
\left\{ {\tau_1 \over \tau_2} + {Q_R\cdot P_R
 \over
 \sqrt{Q_R^2 P_R^2 - (Q_R\cdot P_R)^2}}\right\}\, ,
 \een
  \be \label{edeflambda}
 \Lambda \equiv \beta \, A^{1/2}\, \lambda_2^{1/2} 
 \,  \left[
{1 \over \lambda_2\tau_2 } + {|\tau|^2|\lambda|^2\over 
\lambda_2\tau_2} + 2{\tau_1 
\lambda_1 \over
\lambda_2\tau_2} 
+ 2\right]^{-1/2} \, .
 \ee
 In going from the first to the second line of \refb{epath2}
 we have used the expression for $\delta m(Q,P)$ given in
 \refb{ede1}.  We shall take $\beta A^{1/2}$ large
but finite so as to provide sufficient exponential suppression factor
and make the trace finite.
Defining 
\be \label{edefrsv}
 \rho\equiv \rho_1+i\rho_2, \quad \sigma\equiv \sigma_1+i\sigma_2,
 \quad v\equiv v_1+iv_2\, ,
 \ee
we can express \refb{epath2} as
\be \label{epath}
 Tr\left[ (-1)^F
 \exp\left\{ 2\pi i \left( \rho  {P^2\over 2} + \sigma  {Q^2\over 2}
 + v \, Q\cdot P \right) \right\}\right]\, .
 \ee
 Thus we see that the path integral automatically
 leads to the dyon partition function with complex
 $(\rho,\sigma, v)$, with the imaginary parts of $(\rho,\sigma,v)$
 given by the prescription 
 of \cite{0706.2363} given in \refb{eimdefpre}. Due to the
 $(-1)^F$ insertion in the part integral the trace will include
 sum over BPS states only.
 We note however that in order to relate \refb{epath}
 to the dyon partition
 function defined in \refb{ef3} we must insert additional 
 projection operators into the trace in \refb{epath}
 which restrict the states
 over which we sum. This is because a given set of values of
 $(Q^2,P^2,Q\cdot P)$ may arise from many different charge
 vectors and in defining the partition function \refb{ef3} we
 have summed over distinct values of $(Q^2,P^2,Q\cdot P)$
 instead of over all charges. For example \refb{e1} shows
 that $Q\cdot P$ depends only on the combination $m_2+n_1$;
 thus for fixed $Q^2$, $P^2$
 if we want to count each value of $Q\cdot P$ only once we must
 not perform independent sums over $m_2$ and $n_1$.
 We have also assumed implicitly that we are summing over
 states which do not carry any charges associated with a 3-brane
 wrapped on a 2-cycle of $K3$ and a 1-cycle of $S^1\times \wt S^1$.
 Finally, we have restricted the sum over states to the sector
 with zero spatial momentum both along the circles
 $S^1$ and $\wt S^1$ and along the non-compact directions; 
 this requires insertion of yet more
 projection operators.
 We shall not keep track of these projection operators in subsequent
 analysis; these will be important in explicit computation of 
 the partition
 function but not in finding the physical interpretation
 of $(\rho,\sigma,v)$. 
 We shall return to this issue  in \S\ref{subtle}.

 In \S\ref{sgeometry} we shall find a geometric description of this
 partition function following the duality maps given in
 \cite{0506249,0612011}.
 
 \sectiono{Chemical Potentials to Period Matrix} \label{sgeometry}
 
In this section we shall find a geometric interpretation of the
dyon partition function introduced in the previous section. 
For this we need to use a dual M-theory description of the
theory.
We first 
make a T-duality transformation along the euclidean time
circle
 to map the  euclidean type IIB string theory described in the
 previous section
 to euclidean type IIA string theory compactified on
 $K3\times S^1\times \wt S^1\times S^1_T$, --
$S^1_T$ being
the circle dual to the euclidean time circle of IIB.  This in turn
 can now be regarded as  euclidean M-theory on 
 $K3\times S^1\times \wt S^1\times S^1_T\times S^1_M$.
 Following the chain of dualities we can find the relationship between
 the parameters labelling the M-theory torus
 $T^4\equiv 
 S^1\times \wt S^1\times S^1_T\times S^1_M$ and the
 parameters labelling the original type IIB compactification.
 In particular one finds that  the parameters $\tau$, $\lambda$,
 $C^{(2)}_{\tau y}$,  $C^{(2)}_{\tau \wt y}$,
 $B^{(2)}_{\tau y}$ and  $B^{(2)}_{\tau \wt y}$
 of the type IIB
 string theory can be regarded as components of the metric
 along
 the four torus in the M-theory description.
Alternatively we can represent the
 M-theory torus as euclidean space with a standard metric
 \be \label{e31}
 (dx^1)^2 + (dx^2)^2 + (dy^1)^2 + (dy^2)^2 \, ,
 \ee
 modded out by a lattice $\Lambda$. 
 In this case the information about 
 $\tau$, $\lambda$,
 $C^{(2)}_{\tau y}$,  $C^{(2)}_{\tau \wt y}$,
 $B^{(2)}_{\tau y}$ and  $B^{(2)}_{\tau \wt y}$ is encoded in the
 generators of the lattice $\Lambda$.
 
 First consider the case when all the off-diagonal
 fields, {\it e.g.} $\tau_1$, $\lambda_1$, $C^{(2)}_{ij}$ and
 $B^{(2)}_{ij}$ vanish. 
 In this case $T^4$ is a direct product of four circles. 
 Let 
 $x^1$, $x^2$, $y^1$ and $y^2$ denote
 coordinates along the circles $S^1_M$,
 $S^1_T$, $\wt S^1$ and $S^1$ respectively. 
 Standard duality transformation rules then tell us that these circles
 have periods $L_2$, $L_2\lambda_2$, $L_1$ and $L_1\tau_2$ 
 respectively for 
 \be \label{eparam}
 L_1 = 2\pi\, \beta^{1/3} \lambda_2^{1/3} A^{1/2} \tau_2^{-1/2},
 \qquad L_2 = 2\pi\, \lambda_2^{-2/3}\, \beta^{-2/3}\, .
 \ee
  This is associated with a lattice $\Lambda$ generated by the
 unit vectors
 \be \label{elead}
\pmatrix{L_2\cr 0\cr 0\cr 0}, \qquad  
 \pmatrix{0\cr
 L_2\lambda_2\cr 0\cr 0}, \qquad
\pmatrix{ 0 \cr  0 \cr L_1\cr 0}, \qquad
\pmatrix{ 0 \cr  0 \cr 0  \cr L_1 \tau_2}\, .
 \ee
 Furthermore the
 volume of $K3$ in the final M-theory metric is related to that
 measured in the original $IIB$ metric via the relation
 \be \label{evol}
 V^M_{K3} = \beta^{4/3} \, \lambda_2^{4/3} \, V_{K3} 
 \,  .
 \ee
 
 As mentioned above, 
 deformations associated with the parameters 
 $\tau_1$, $\lambda_1$, 
 $C^{(2)}_{ij}$ and
 $B^{(2)}_{ij}$
 can be shown to be associated with the geometric deformation of
 the four torus  spanned by $S^1$, $\wt S^1$, $S^1_M$ and
 $S^1_T$. One can find the parameters of the deformed torus
 in terms of the parameters of the type IIB theory.
The four generators
 of the lattice $\Lambda$ associated with this deformed
 M-theory torus
 turn out to be
 \be \label{e32}
 e_1 = \pmatrix{L_2\cr 0\cr 0\cr 0}, \quad e_2=
 \pmatrix{-L_2\lambda_1\cr
 L_2\lambda_2\cr 0\cr 0}, \quad
 e_3 = \pmatrix{ L_2 C^{(2)}_{\tau \wt y} - L_2 \lambda_1
  B^{(2)}_{\tau
 \wt y}\cr  L_2\lambda_2 B^{(2)}_{\tau
 \wt y} \cr L_1\cr 0}, \quad
 e_4 = \pmatrix{ L_2 C^{(2)}_{\tau y} - L_2 \lambda_1 
 B^{(2)}_{\tau
  y} \cr  L_2 \lambda_2 B^{(2)}_{\tau
 y} \cr -L_1\tau_1 \cr L_1 \tau_2}\, .
 \ee
 Equivalently one can describe the M-theory torus as a product
 of four circles, each of period $2\pi$, with metric
 $g_{ij}=e_i\cdot e_j$. Note that a shift of $C^{(2)}_{\tau y}$,
 $C^{(2)}_{\tau \wt y}$, $B^{(2)}_{\tau y}$ or
 $B^{(2)}_{\tau \wt y}$ by an integer produces an integer linear
 combination of the original $e_i$'s and  generate the same
 lattice.
  
 Our next task will be to study the fate of the original D5-NS5-brane
 configuration in the M-theory description. First consider the case
 where $\lambda_1$, $\tau_1$, $C^{(2)}_{ij}$ and
 $B^{(2)}_{ij}$ vanish. In this case the D5-brane wrapped along
 $K3\times \wt S^1$ becomes an euclidean
 M5-brane wrapped along $K3\times \wt S^1 \times S^1_M$ and the
 NS5-brane wrapped on $K3\times  S^1$ becomes an euclidean
 M5-brane wrapped on $K3\times  S^1\times S^1_T$.
 Leaving aside the K3 part, this can be regarded as a degenerate
 genus two surface embedded in $T^4$, with period matrix
 \be \label{e33}
 \Omega \equiv \pmatrix{\ks & \kv\cr \kv & \krh}
 = \pmatrix{i\, {L_1 / L_2} & 0\cr
 0 &  i\, L_1 \tau_2/ ( L_2 \lambda_2)}\, .
 \ee
 For more general background in the original type IIB string theory,
 the M5-brane configuration in the dual M-theory will have the
 form of $K3$ times a non-singular genus two surface embedded
 in $T^4$. Our goal now will be to determine the period matrix
 of this genus two surface in terms of the parameters labelling the
 original type IIB background. The main tool will be supersymmetry
 which requires that the genus two surface is holomorphically
 embedded in $T^4$. Let $\omega_1$ and $\omega_2$ be the
 two linearly independent holomorphic 1-forms on the genus two
 surface and let $A_1$, $A_2$, $B_1$, $B_2$ be a basis of integral
 homology cycles satisfying
 \be \label{e34}
 A_i\cap B_j = \delta_{ij}, \quad A_i\cap A_j=B_i\cap B_j=0\, .
 \ee
 Then the requirement that the genus two surface is holomorphically
 embedded in $T^4$ amounts to the constraint that the $T^4$ is
 given by $\CCC^2/\overline{\Lambda}$, where the lattice
 $\overline{\Lambda}$ is generated by the 
 complex two dimensional vectors\cite{reference}
 \be \label{e35}
 \bar e_1 = \pmatrix{\ointop_{A_1} \omega_1\cr 
 \ointop_{A_1} \omega_2}, \quad
 \bar e_2 = \pmatrix{\ointop_{A_2} \omega_1\cr 
 \ointop_{A_2} \omega_2}, \quad
\bar e_3= \pmatrix{\ointop_{B_1} \omega_1\cr 
 \ointop_{B_1} \omega_2}, \quad
\bar e_4= \pmatrix{\ointop_{B_2} \omega_1\cr 
 \ointop_{B_2} \omega_2}\, .
 \ee
 $\CCC^{(2)}$ is endowed with the standard metric $|dz_1|^2+
 |dz_2|^2$. Now by taking appropriate complex linear combinations
 of the two $\omega_i$'s we can guarantee that
 \be \label{e36}
 \ointop_{A_i}\omega_j = \delta_{ij}, \quad
 \int_{B_i} \omega_j = \Omega_{ij}\, ,
 \ee
 where $\Omega=\pmatrix{\ks & \kv\cr \kv &\krh}$ 
 is the period matrix of the genus two surface.
 Thus for a general choice of the $\omega_i$'s, related to the one above
 by a $GL(2,\CCC)$ transformation matrix $V$, we have
 \be \label{e37}
 \bar e_1 = V \pmatrix{1\cr 0}, \quad \bar e_2 = V\pmatrix{0\cr 1},
 \quad \bar e_3 = V\pmatrix{\ks\cr \kv}, \quad \bar e_4
 = V \pmatrix{\kv\cr\krh}\, .
 \ee
 
 Our goal is to compare, up to $SO(4)$ rotations, 
 the basis vectors $\bar e_i$ given in
 \refb{e37} with those given in \refb{e32} to find the relationship
 between the parameters $\tau$, $\lambda$, 
 $C^{(2)}_{ij}$, $B^{(2)}_{ij}$ of the M-theory torus and the
 parameters $\krh$, $\ks$ and $\kv$ labelling the moduli
 of the genus two surface. For this we have to express the
 two component
 complex vectors $\bar e_i$
 given in \refb{e37} as four component real
 vectors by separately picking out the
 real and imaginary parts of $\bar e_i$. 
 If $\bar e_{iR}$ and $\bar e_{iI}$
 denote the real and the imaginary parts of $\bar e_i$, then the
 inner product between $\bar e_i$ and $\bar e_j$, regarded as
 real vectors, is given by 
 $\bar e_{iR}\cdot \bar e_{jR}+\bar e_{iI}\cdot \bar e_{jI}
 = Re(\bar e_i^*\cdot \bar e_j)$.
 Defining $a$, $b$, $c$ through
 \be \label{e38}
 V^\dagger V 
 = \pmatrix{a & c \cr c^* & b}, \quad a,b\in \RRR, \quad c\in \CCC\, ,
 \ee
 we get
 \be \label{e39}
 Re(\bar e_i^*\cdot \bar e_j) = Re\pmatrix{
 a & c & a\ks+c\kv &  a\kv+c\krh\cr
 c^* & b & c^*\ks+b\kv &  c^* \kv + b\krh \cr
 \ks^* a + \kv^* c^* & \ks^* c + \kv^* b & \ks^*(a\ks+c\kv)
  & \ks^* ( a\kv+c\krh)\cr
 & & + \kv^*(c^*\ks+b\kv) & +\kv^*(c^*\kv+b\krh) \cr
  \kv^* a + \krh^* c^* &  \kv^*c + \krh^* b & 
  \kv^*(a\ks+c\kv) &
  \kv^* ( a\kv+c\krh) \cr
 & & +\krh^*(c^*\ks+b\kv) & + \krh^*( c^* \kv+b\krh)}\, .
 \ee
 We can now compare this with 
 the inner product matrix $e_i\cdot e_j$
 constructed from 
 the basis vectors given in \refb{e32}.  
 In computing $e_i\cdot e_j$  we 
 shall consider the special case
 where $C^{(2)}_{\tau  y}=B^{(2)}_{\tau \wt y}$, since
 this is the background used in the analysis of \S\ref{sfivebrane}. 
 In this case
 the equations $Re(\bar e_i^*\cdot \bar e_j)=e_i\cdot e_j$ 
 can be solved to give:
 \ben \label{e40}
 && a = L_2^2, \quad b = L_2^2 |\lambda|^2, \quad c 
 = -L_2^2 \lambda_1, 
 \quad \ks_1=C^{(2)}_{\tau \wt y}, \quad 
 \krh_1=B^{(2)}_{\tau  y},
 \quad \kv_1 = C^{(2)}_{\tau y} = 
 B^{(2)}_{\tau \wt y}, \nonumber \\
 && \ks_2^2 - 2\lambda_1 \kv_2\ks_2 + 
 |\lambda|^2 \kv_2^2 = L_1^2/L_2^2\, ,
 \nonumber \\
 &&  \ks_2 \kv_2 - \lambda_1 
 (\ks_2\krh_2 + \kv_2^2) + |\lambda|^2
 \kv_2 \krh_2 = -\tau_1 L_1^2 / L_2^2\, , \nonumber \\
 && \kv_2^2 - 2\lambda_1 \kv_2 \krh_2 
 + |\lambda|^2 \krh_2^2 = |\tau|^2
 L_1^2 / L_2^2 \, .
 \een
 The equations for $\ks_2$, $\krh_2$ and $\kv_2$ can be solved
 as
 \ben \label{e41}
 \ks_2 &=& \check\Lambda \left\{ {|\lambda|^2 \over \lambda_2} 
 + {1\over \tau_2}
 \right\} = \check\Lambda \left\{ {1\over \tau_2}
 + {P_{0R}^2\over
 \sqrt{Q_{0R}^2 P_{0R}^2 - (Q_{0R}\cdot P_{0R})^2}}\right\}\, , \nonumber \\
\krh_2 &=& \check\Lambda \left\{ {1 \over \lambda_2} + {|\tau|^2
\over \tau_2}
 \right\} = \check\Lambda \left\{ {|\tau|^2
\over \tau_2} + {Q_{0R}^2\over
 \sqrt{Q_{0R}^2 P_{0R}^2 - (Q_{0R}\cdot P_{0R})^2}}\right\}\, , \nonumber \\
 \kv_2 &=& \check\Lambda 
 \left\{ {\lambda_1 \over \lambda_2} - {\tau_1
\over \tau_2}
 \right\} = -\check\Lambda \left\{ {\tau_1
\over \tau_2}
 + {Q_{0R}\cdot P_{0R}
 \over
 \sqrt{Q_{0R}^2 P_{0R}^2 - (Q_{0R}\cdot P_{0R})^2}}\right\}\, ,
 \een
 where we have used eqs.\refb{e22} and
 \be \label{e42}
\check \Lambda = L_1 L_2^{-1} \, \tau_2\,
\left\{ 1+2\lambda_1\tau_1 +2\lambda_2\tau_2
+|\lambda|^2 |\tau|^2 \right\}^{-1/2}
 \, .
 \ee

 Finally let us discuss the effect of inclusion of the D-strings and
 fundamental strings in the original type IIB description. These
 correspond to appropriate excitations on the M5-brane world
 volume. However in the $V_{K3}\to\infty$ limit we expect the
 effect of these excitations on the M5-brane geometry to vanish
 and $(\krh,\ks,\kv)$ given by \refb{e40}, 
 \refb{e41} continue to describe
 the moduli of the genus two surface which the M5-brane wraps.
 Furthermore in
 this limit we can replace $(Q_{0R}^2, P_{0R}^2, Q_{0R}\cdot
 P_{0R})$ by $(Q_R^2, P_R^2, Q_R\cdot P_R)$ in \refb{e41}.
 Comparing \refb{e40}, \refb{e41}, \refb{e42} with
 \refb{en1}, \refb{eimdef}, \refb{edeflambda}, 
 we now see that
 \be \label{e51}
 \check\Lambda = \Lambda\, ,
 \ee
 and
 \be \label{e52}
 \krh=\rho, \quad \ks=\sigma, \quad
 \kv=v\, .
 \ee
 This shows that the dyon partition function given in
 \refb{epath} is given by an appropriate partition function on
 a genus two Riemann surface with modular parameters
 $(\rho,\sigma,v)$. Furthermore, as already noted below
 \refb{epath}, the imaginary parts of $(\rho,\sigma, v)$
 are automatically set according to the prescription 
given in
 \cite{0706.2363}.
 
 \sectiono{The Partition Function} \label{subtle}
 
 While our analysis determines the moduli of the genus two
 surface on which the M5-brane is wrapped, we have not
 determined
 precisely what computation we need to perform on this genus
 two surface to extract the partition function. Since the low
 energy world-sheet
 theory of
 M5-brane wrapped on K3 coincides with that of 
 a fundamental heterotic
 string on a transverse
 $T^3$ in static gauge Green-Schwarz
 formulation, one might expect the final partition function to be
 given by that of an euclidean 
 heterotic string in 
 $T^4\times T^3$\cite{0506249,0612011}.\footnote{Note that
in the $V^M_{K3}\to\infty$ limit, describing a K3 wrapped
M5-brane as a fundamental heterotic string as in
\cite{0506249,0612011} is not useful in general. However
for computing an index one may still be able to 
use this description.}
 However such a partition function vanishes due to
the right-moving fermion zero modes, 
and one must pick only the left-moving
 part of the partition function to get the desired result given by the
 inverse of the Igusa cusp form. One might expect that this
 prescription should follow from the fact that in defining the
 original partition function in the type IIB string theory we have
 removed by hand the right-moving fermion zero modes,
but exactly what this translates to in the M-theory
description is not 
 understood\cite{0612011}. One possibility will be to begin
 with the helicity trace by inserting six powers of helicity in the
 original trace in type IIB string theory and then carefully keeping
 track of this factor during the duality transformations.
  
 Another issue arises from the need to insert additional
 projection operators into the trace in the original type IIB
 description so that the trace  receives contribution only
 from a subset of dyon charges. As mentioned below \refb{epath},
 this is necessary to ensure that in the trace each value of
 $(Q^2, P^2,Q\cdot P)$ is counted only once. 
 In particular although the full set of allowed charges consists
 of 28 electric charges and 28 magnetic charges, the charge
 configuration we have considered span only a small subspace.
 Furthermore we have restricted the trace to be over states carrying
 zero spatial momenta.
 What do such restrictions correspond to in the M-theory picture?
 To answer this question we first note that in the original type IIB
 description the missing charges
 can be divided into four classes: 1) Kaluza-Klein monopoles
 associated with the circles $S^1$ and $\wt S^1$, 
 2) more general 5-branes
 wrapped on $K3$ times $S^1$ or $\wt S^1$, 3) 
 charges associated
 with D3-branes wrapped on a 2-cycle of $K3$ and $S^1$ 
 or $\wt S^1$ and 4) 
 momenta
 along  $S^1$ and $\wt S^1$. 
 Under the duality map of \S\ref{sgeometry} 
 that takes the euclidean
 type IIB string theory to the euclidean M-theory, the Kaluza-Klein
 monopoles get mapped to Kaluza-Klein monopoles. 
 Since we do not expect the partition function associated with an
 euclidean M5-brane to include contribution from the Kaluza-Klein
 monopole sector, it is natural to set the Kaluza-Klein monopole
 charges in the original type IIB theory to zero.
On the other hand a configuration carrying
 a  general set of 5-brane charges compared to the one considered
 in \S\ref{sfivebrane}, {\it e.g.} a $(p_1,q_1)$ 5-brane along
 $K3\times \wt S^1$ and a $(p_2,q_2)$ 5-brane along 
 $K3\times S^1$,
will get mapped to a
euclidean M5-brane configuration wrapping $K3$ times a genus
2-surface in $T^4$, but the
embedding of the genus two surface into $T^4$ will be topologically
distinct for different $(p_1,q_1,p_2,q_2)$. Thus the M5-brane
partition function, regarded as a partition function on a genus
two surface, will naturally 
include contribution from states with a fixed set
of  $(p_1,q_1,p_2,q_2)$ which in our case is
$(1,0,0,1)$. 
Charges associated
 with D3-branes wrapped on a 2-cycle of $K3$ and $S^1$ 
 or $\wt S^1$ correspond to switching on magnetic flux
 on the D5 or NS5-brane world-volume 
 along a 2-cycle of K3. By following the
 chain of dualities one finds that this corresponds to switching on
 the flux of the 3-form field strength on the euclidean M5-brane
 along a 2-cycle of K3 times $S^1_M$ or $S^1_T$. Thus requiring that
 the D3-brane charges vanish in the original type IIB theory amounts
 to restricting the path integral over M5-brane degrees of freedom to
 sectors with zero 3-form field strength flux through the 2-cycles of
 $K3$ times $S^1_M$ or $S^1_T$.\footnote{In our convention
 this corresponds to zero 3-form field strength along the
 product of a 2-cycle on K3 and one of the A-cycles on the genus
 two surface. Since the 3-form field strength is self-dual, this
 effectively restricts all possible flux of the 3-form field strength
 on the euclidean M5-brane world-volume to zero
 barring the issue that in an euclidean signature space the 
 requirement
 of a self-duality makes a 3-form complex.}
 Finally the effect of  switching
 on momenta, either along $S^1$ or $\wt S^1$ or along the non-compact
 directions, corresponds to collective motion of the string network and
 does not affect either the degeneracy of states or the values of $Q^2$,
 $P^2$ and $Q\cdot P$. Thus on the type IIB side
 restricting these momenta to zero should correspond to factoring
 out (or freezing) the euclidean path integral over these
 collective modes. We need to determine what
 operation this corresponds to on the
 M5-brane partition function; we shall come back to this
 issue shortly.
 
This does not exhaust the list of restrictions 
we need to impose on the
M5-brane path integral.
As mentioned below
\refb{epath}, independent  sum over $m_2$ and $n_1$ gives
the same $(Q^2,P^2,Q\cdot P)$ infinite number of times
since the latter depends on $m_2$ and $n_1$ only through the
combination $m_2+n_1$. Thus charge vectors of the form
$(n_1+k, m_2-k)$ generate the same set of invariants as
$(n_1,m_2)$. We shall now argue that this sum over $k$ can also
be regarded as a sum over momentum conjugate to a collective
mode. For this we note that if we just have a D5-brane wrapped on
$K3\times \wt S^1$, then we can generate a fundamental string charge
along $\wt S^1$ by switching on an electric field along $\wt S^1$.
This can be interpreted as the momentum conjugate to the
Wilson line along $\wt S^1$ on the D5-brane, and contributes to
the quantum number $n_1$. Similarly by switching on an electric
field on NS 5-brane along $K3\times S^1$ we can generate a
D-string charge along $S^1$. This is the momentum conjugate to
the Wilson line along $S^1$ on the NS 5-brane and contributes
to $-m_2$. However once the D5-brane and the NS 5-brane join to
form a string network as in Fig.~\ref{f1},
the Wilson lines on D5-brane and NS 5-branes
do not describe independent collective coordinates. To see this
we note from Fig.~\ref{f1}
that in the string network configuration there is an
intermediate $(1,1)$ 5-brane, and switching on an electric field along
this generates equal but opposite  
amount of D-string and the 
fundamental string charge. Thus by requiring conservation  of 
fundamental and D-string charges at the junction we see that the
electric fields carried by the D5-brane and the NS 5-brane must
be correlated such that the net D-string charge carried along $S^1$
and the net fundamental string charge carried along $\wt S^1$ are
equal. In other words only changes of the form $(n_1,m_2)\to
(n_1+k,m_2-k)$ can be regarded as due to excitations of a
collective coordinate. 
Thus picking one representative $(m_2,n_1)$
for each $m_2+n_1$ corresponds to freezing this collective degree
of freedom. 
On the other hand fluctuations of the string network
which change the values of $m_2+n_1$ cannot be regarded as
collective coordinate excitations.

Thus our task now is to determine what operation on the
M-theory side would correspond to freezing or factoring out
the contribution from the various collective modes of the string
network on the type IIB side.
By following the chain of dualities leading from the IIB description
to the M-theory description we can identify the collective 
deformations
of the  network in the euclidean
IIB theory to those of the M5-brane
in euclidean M-theory. In particular the collective 
deformations associated with the
translation of the network along the non-compact directions, $S^1$
and $\wt S^1$ correspond respectively to the freedom of
translating the euclidean M5-brane along the 
non-compact directions, $S^1$
and $\wt S^1$.
On the other hand
the collective deformation corresponding to the Wilson line on the
network corresponds to the freedom of switching on an
anti-symmetric 2-form field, proportional to the volume form
on the genus two surface, on the M5-brane world-volume. 
Nevertheless it is not entirely clear what a 
time dependent collective
excitation of the string junction 
in IIB side translates to in the M-theory side. 
If $\phi$ denotes a collective mode of the string network
then freezing it would correspond to setting $\phi(\tau)=0$
in the euclidean path integral. While the zero mode of $\phi(\tau)$
has a simple interpretation in the dual description involving the
euclidean M5-brane, the physical interpretation of the
$\tau$ dependent part of $\phi(\tau)$ in the M5-brane
world-volume theory is more complicated
since
the duality relating the two descriptions involves  a T-duality
transformation that converts momentum modes along the
euclidean time circle into fundamental string
winding modes and vice versa.

To summarize, in order that the partition function 
of the string network
in the 
type IIB description computes the dyon partition function of our
interest, we must freeze
all the collective excitations of the
string network
 (or factor out their contribution from the partition function), 
 and at the same time set the magnetic flux on the
5-branes through the 2-cycles of K3 to zero. In the M-theory
description the latter can be
interpreted as setting to zero appropriate flux of 3-form field strength
on the M5-brane. However the physical interpretation of freezing
the collective coordinates of the string network is not entirely
clear since the duality relating the type IIB description and the
M-theory description involves T-duality along the euclidean
time circle.
 One could however be a little less ambitious and take the
 point of view that since on the type IIB side the contribution from
 the bosonic (fermionic)
 collective modes do not affect the $(\rho,\sigma,v)$ dependence
 of the partition function but only generate an overall 
 divergent (zero) factor, we
 could begin with the partition function of the euclidean string
 network without freezing any collective modes and then
 simply throw away  the overall $(\rho,\sigma,v)$ 
 independent
 divergent or zero factors. This would translate to a
 similar operation on the M5-brane partition function.
 This clearly would not determine the overall normalization of the
 partition function but could determine its dependence  
 on $(\rho,\sigma,v)$.
 Along this line one could also make use of the fact
 that in the type IIB description
 the partition function $Z$ depends only on $(\rho,\sigma, v)$ and
 not their complex conjugates due to \refb{epath}. Thus in the final
 answer we can just pick up the holomorphic part of the
 M5-brane partition function; any dependence on
 $(\bar\rho,\bar\sigma,\bar v)$  must
 cancel at the end.

 If we accept the above proposal then the computation of the
 partition function will involve computing the M5-brane partition
 function with certain restriction on the 3-form fluxes and
 picking up $(\rho,\sigma,v)$ dependence of the result without
 worrying about any overall divergence or zero coming from
 bosonic or fermionic zero mode integration. Even then the
 analysis is not entirely straightforward since the world-volume theory
 of the wrapped M5-brane resembles that of 
 heterotic string in static gauge
 Green-Schwarz formulation and most of the explicit genus
 two computations in heterotic string theory are performed in the
 Neveu-Schwarz-Ramond formulation.
We hope to return to these issues in the near future.

We end by noting that knowing the partition function 
 is important not only for finding the explicit
 form of the dyon spectrum but also for understanding why as
 an analytic function the partition function retains its form as we
 cross a wall of marginal stability.

\noindent{\bf Acknowledgement:} 
We would like to thank Miranda Cheng and Atish Dabholkar for useful
discussions. We would also like to thank Miranda Cheng for
her critical
comments on an earlier version of the manuscript which helped
us fix a sign error.
The research of A.S. was supported
in part by the JC Bose fellowship. We would also like to
acknowledge the hospitality of the `Monsoon workshop on string theory'
at TIFR where part of the work was done.



\begin{thebibliography}{99}


\bibitem{9607026}
R.~Dijkgraaf, E.~P.~Verlinde and H.~L.~Verlinde,
``Counting dyons in N = 4 string theory,''
Nucl.\ Phys.\ B {\bf 484}, 543 (1997)
[arXiv:hep-th/9607026].

\bibitem{0412287}
  G.~Lopes Cardoso, B.~de Wit, J.~Kappeli and T.~Mohaupt,
  ``Asymptotic degeneracy of dyonic 
  N = 4 string states and black hole
  entropy,''
  JHEP {\bf 0412}, 075 (2004)
  [arXiv:hep-th/0412287].

\bibitem{0505094}
D.~Shih, A.~Strominger and X.~Yin,
``Recounting dyons in N = 4 string theory,''
arXiv:hep-th/0505094.

\bibitem{0506249}
D.~Gaiotto,
``Re-recounting dyons in N = 4 string theory,''
arXiv:hep-th/0506249.

\bibitem{0508174}
  D.~Shih and X.~Yin,
  ``Exact black hole degeneracies and the topological string,''
  JHEP {\bf 0604}, 034 (2006)
  [arXiv:hep-th/0508174].

\bibitem{0510147}
  D.~P.~Jatkar and A.~Sen,
  ``Dyon spectrum in CHL models,''
  JHEP {\bf 0604}, 018 (2006)
  [arXiv:hep-th/0510147].

\bibitem{0602254}
  J.~R.~David, D.~P.~Jatkar and A.~Sen,
  ``Product representation of dyon partition function in CHL models,''
  JHEP {\bf 0606}, 064 (2006)
  [arXiv:hep-th/0602254].
  
\bibitem{0603066}
  A.~Dabholkar and S.~Nampuri,  
  ``Spectrum of dyons and black holes in 
  CHL orbifolds using Borcherds lift,''
  arXiv:hep-th/0603066.

\bibitem{0605210}
  J.~R.~David and A.~Sen,
  ``CHL dyons and statistical entropy function from D1-D5 system,''
  JHEP {\bf 0611}, 072 (2006)
  [arXiv:hep-th/0605210].

\bibitem{0607155}
  J.~R.~David, D.~P.~Jatkar and A.~Sen,
  ``Dyon spectrum in N = 4 supersymmetric type II string theories,''
  arXiv:hep-th/0607155.


\bibitem{0609109}
  J.~R.~David, D.~P.~Jatkar and A.~Sen,
  ``Dyon spectrum in generic N = 4 supersymmetric Z(N) orbifolds,''
  arXiv:hep-th/0609109.

\bibitem{0612011}
  A.~Dabholkar and D.~Gaiotto,
  ``Spectrum of CHL dyons from genus-two partition function,''
  arXiv:hep-th/0612011.

\bibitem{0702141}
  A.~Sen,
  ``Walls of marginal stability and dyon spectrum in N = 4 supersymmetric
  string theories,''
  arXiv:hep-th/0702141.

\bibitem{0702150}
  A.~Dabholkar, D.~Gaiotto and S.~Nampuri,
  ``Comments on the spectrum of CHL dyons,''
  arXiv:hep-th/0702150.
  

\bibitem{0705.1433}
  N.~Banerjee, D.~P.~Jatkar and A.~Sen,
  ``Adding charges to N = 4 dyons,''
  arXiv:0705.1433 [hep-th].

\bibitem{0705.3874}
  A.~Sen,
  ``Two Centered Black Holes and N=4 Dyon Spectrum,''
  arXiv:0705.3874 [hep-th].
    
\bibitem{0706.2363}
  M.~C.~N.~Cheng and E.~Verlinde,
  ``Dying Dyons Don't Count,''
  arXiv:0706.2363 [hep-th].

\bibitem{0707.1563}
  A.~Sen,
  ``Rare Decay Modes of Quarter BPS Dyons,''
  arXiv:0707.1563 [hep-th].

\bibitem{0707.3035}
  A.~Mukherjee, S.~Mukhi and R.~Nigam,
  ``Dyon Death Eaters,''
  arXiv:0707.3035 [hep-th].

\bibitem{0708.1270}
  A.~Sen,
  ``Black Hole Entropy Function, 
Attractors and Precision Counting of
  Microstates,''
  arXiv:0708.1270 [hep-th].

 \bibitem{0710.4533}
  A.~Mukherjee, S.~Mukhi and R.~Nigam,
  ``Kinematical Analogy for Marginal Dyon Decay,''
  arXiv:0710.4533 [hep-th].

\bibitem{0712.0043}
S.~Banerjee and A.~Sen, 
``Duality Orbits, Dyon Spectrum and Gauge Theory Limit of
Heterotic String Theory on $T^6$'',
arXiv:0712.0043 [hep-th].
  
\bibitem{0801.0149}
  S.~Banerjee and A.~Sen,
  ``S-duality Action on Discrete T-duality Invariants,''
  arXiv:0801.0149 [hep-th].

\bibitem{0802.0544}
  S.~Banerjee, A.~Sen and Y.~K.~Srivastava,
  ``Generalities of Quarter BPS Dyon Partition Function and Dyons of Torsion
  Two,''
  JHEP {\bf 0805}, 101 (2008)
  [arXiv:0802.0544 [hep-th]].

\bibitem{0802.1556}
  S.~Banerjee, A.~Sen and Y.~K.~Srivastava,
  ``Partition Functions of Torsion >1 Dyons in Heterotic String Theory on
  $T^6$,''
  JHEP {\bf 0805}, 098 (2008)
  [arXiv:0802.1556 [hep-th]].

\bibitem{0803.2692}
  A.~Dabholkar, J.~Gomes and S.~Murthy,
  ``Counting all dyons in N =4 string theory,''
  arXiv:0803.2692 [hep-th].
 
\bibitem{0803.3857}
  A.~Sen,
  ``Wall Crossing Formula for N=4 Dyons: A Macroscopic Derivation,''
  arXiv:0803.3857 [hep-th].

\bibitem{0806.2337}
  M.~C.~N.~Cheng and E.~P.~Verlinde,
  ``Wall Crossing, Discrete Attractor Flow, and Borcherds Algebra,''
  arXiv:0806.2337 [hep-th].

\bibitem{9704170}
  O.~Aharony and A.~Hanany,
  ``Branes, superpotentials and superconformal fixed points,''
  Nucl.\ Phys.\  B {\bf 504}, 239 (1997)
  [arXiv:hep-th/9704170].

\bibitem{9710116}
  O.~Aharony, A.~Hanany and B.~Kol,
  ``Webs of (p,q) 5-branes, five dimensional field theories and grid
  diagrams,''
  JHEP {\bf 9801}, 002 (1998)
  [arXiv:hep-th/9710116].

\bibitem{9711130}
  A.~Sen,
  ``String network,''
  JHEP {\bf 9803}, 005 (1998)
  [arXiv:hep-th/9711130].

\bibitem{0005049}
  F.~Denef,
  ``Supergravity flows and D-brane stability,''
  JHEP {\bf 0008}, 050 (2000)
  [arXiv:hep-th/0005049].

\bibitem{0010222}
F.~Denef,
``On the correspondence between D-branes 
and stationary supergravity solutions of type
II Calabi-Yau compactifications'', 
arXiv:hep-th/0010222.

\bibitem{0101135}
  F.~Denef, B.~R.~Greene and M.~Raugas,
  ``Split attractor flows and the spectrum 
of BPS D-branes on the 
quintic,''
  JHEP {\bf 0105}, 012 (2001)
  [arXiv:hep-th/0101135].
  
\bibitem{0206072}
  F.~Denef,
  ``Quantum quivers and Hall/hole halos,''
  JHEP {\bf 0210}, 023 (2002)
  [arXiv:hep-th/0206072].

\bibitem{0304094}
  B.~Bates and F.~Denef,
  ``Exact solutions for supersymmetric stationary black hole 
composites,''
  arXiv:hep-th/0304094.

\bibitem{0702146}
  F.~Denef and G.~W.~Moore,
  ``Split states, entropy enigmas, holes and halos,''
  arXiv:hep-th/0702146.

\bibitem{9708062}
  A.~Gregori, E.~Kiritsis, C.~Kounnas, N.~A.~Obers, 
  P.~M.~Petropoulos and B.~Pioline,
  ``R**2 corrections and non-perturbative 
  dualities of N = 4 string ground
  states,''
  Nucl.\ Phys.\ B {\bf 510}, 423 (1998)
  [arXiv:hep-th/9708062].

\bibitem{reference}
P.~Griffiths and J.~Harris,
Principles of Algebraic Geometry, Wiley (1978).

\end{thebibliography}
\end{document}